%% file: Pbreview.tex
\documentclass[twocolumn,showpacs,preprintnumbers,prb,amsmath,amssymb,showkeys,floatfix,nofootinbib,byrevtex]{revtex4}

\usepackage{graphicx}
\usepackage{epsfig} 
\usepackage{dcolumn}
\usepackage{bm}

\begin{document}
\include{defs}

\preprint{Topical Review in the Journal of Physics: Condensed Matter (2004)}

\title{Current-Induced Pair Breaking in Magnesium Diboride}

\author{Milind N. Kunchur} 
 \homepage{http://www.physics.sc.edu/kunchur}
 \email{kunchur@sc.edu}
\affiliation{Department of Physics and Astronomy\\
University of South Carolina, Columbia, SC 29208}

\date{\today}

\begin{abstract}
The transport of electrical current through a
superconductor falls into three broad regimes: non-dissipative,
dissipative but superconducting, and normal or non-superconducting.
These regimes are demarkated by two 
definitions  of critical current: one
is the threshold current above which the superconductor enters a
dissipative (resistive) state; the other is the thermodynamic
threshold above which the superconductivity itself is destroyed and the
superconducting order parameter \del vanishes. 
The first threshold defines the conventional
critical current density \jc and the second defines the 
depairing (or pair-breaking) current \jdy . 
Type II superconductors in the mixed state have
quantized flux vortices, which tend to move when acted upon by the
Lorentz driving force of an applied transport current. In such a mixed
state the resistance vanishes only when vortices are pinned in place by
defects and the applied current is below the threshold \jc required to
overcome pinning and mobilize the vortices. Typically \jdy $\gg$\jc and
a direct experimental measurement of \jd
over the entire temperature range ($0 \leq T \leq$ \tcy ) 
is prohibited by the enormous power dissipation densities
($p \sim 10^{10}$-- $10^{12}$W/\cmcy ) needed to reach the normal state. 
In this work, intense pulsed signals were used to extend transport 
measurements to unprecedented power densities ($p \sim 10^{9}-10^{10}$
W/\cmcy ).
This together with \mgby 's combination of low
normal-state resistivity (\rrhony ) and high transition temperature
(\tcy ) have permitted 
a direct estimation of \jd over the entire temperature range. 
This review describes our experimental investigation 
of current-induced depairing in \mgby ,  and provides an introduction
to the phenomenological theories of superconductivity and how the
observations fit in their context.

\end{abstract}

\maketitle

\section{\label{sec:level1}Introduction}

When a superconductor is cooled below some transition temperature \tcy , 
it undergoes a phase transition leading to the formation of a
superconducting state wherein its charge carriers 
correlate and condense into a coherent macroscopic quantum state. 
In the superconducting state, the system expels magnetic flux upto
magnetic field values below \hcy , the thermodynamic field. In type II
superconductors, the flux expulsion is partial for fields between the
lower and upper critical fields, \hcl and \hcu respectively. The
superconducting state is also characterized by an absence of resistivity
for current densities $j$ below some critical value \jcy. In type II
superconductors, there is partial resistivity for $j$ values between \jc
and \jdy , where \jd is the pair-breaking or depairing current 
density\footnote{Type 
I superconductors can also exhibit partial resistivity in the
superconducting state because of fluctuations, presence of a large
($H>$\hcy ) current induced
self field at the surface (Silsbee's rule), phase slippage as one
approaches the phase boundary, or motion of magnetic domains in the
intermediate state.}.

The formation of this state  
is governed principally by a competition between four energies: condensation,
magnetic-field expulsion, thermal, and kinetic. The order
parameter $\Delta$, that
describes the extent of condensation and the strength of the 
superconducting state, is
reduced as the temperature $T$, magnetic field $H$, and electric current
density $j$ are increased.
The boundary in the $T$-$H$-$j$ phase space that separates the 
superconducting and normal states
is where $\Delta$ vanishes, and the three parameters attain their
critical values $T_{c2}(H,j)$,
$H_{c2}(T,j)$, and $j_{d}(T,H)$. 
$j_{d}$ sets the intrinsic upper limiting scale for 
supercurrent transport in any superconductor. 

Close to \tcy , \jdt has been investigated in several systems
\cite{bardeen,andratskii,skocpol}. In a few materials \cite{romijn,rusanov},
\jdt has been measured down to about 0.2\tcy .
Here we review our investigation of current-induced pair-breaking in
\mgby , which represents a complete ($0 \alt T \alt$ \tcy )
investigation of \jd by a direct transport method.

\section{Theoretical background} 
The magnitude and temperature dependence of the pair-breaking current 
density can be adequately described by
phenomenological theories of superconductivity, such as the London theory
and the Ginzburg-Landau (GL) theory. In a microscopic theory such as the
Bardeen-Stephen-Cooper (BCS) theory, experimental quantities are
calculated from microscopic parameters such as the strength of the
effective attractive interaction that leads to Cooper pair formation and
the density of states at the Fermi level. Often these microscopic
parameters are not sufficiently well known. In phenomenological
theories, connections are made between the different observables from
contraints based on thermodynamic principles and
electrodynamical properties of the superconducting state. Below we
give a simplified introduction to the different theoretical approaches for
estimating the depairing current density and its \tdepy .

\subsection{London theory}

\subsubsection{Basic framework}

The London theory \cite{london} 
of superconductivity provides a description of the observed
electrodynamical properties by supplementing the basic Maxwell equations
by additional equations, which contrain the possible behavior to reflect
the two hallmarks of the superconducting state---perfect conductivity and
Meissner effect. Note that these properties hold only partially
when mobile vortices are present. 

An ordinary metal (normal conductor) requires a driving electric field $E$
to maintain a constant current against resistive losses. 
In the simple Drude picture, this produces Ohm's
law behavior, $j=\sigma E$, with a conductivity given by 
$\sigma=ne^{*2}\tau/m^{*}$,
where $n$ is the number density of charge carriers, and 
$e^{*}$ and $m^{*}$  are their effective charge and mass respectively. 

A superconductor can carry a resistanceless
current and so an electric field is not required to carry a persistent current.
Instead $E$ in a perfectly conducting state causes a ballistic
acceleration of charge so that $\partial j/\partial t \propto E$.
If the number density of effective charge carriers (pairs) 
participating in the supercurrent is $n^{*}_{s}$, 
then $j=n^{*}_{s}e^{*}v_{s}$, where $v_{s}$ is the carrier (superfluid) 
velocity, and
\be \label{london1}
E=\left( \frac{m^{*}}{n^{*}_{s}e^{*2}}\right) \frac{\partial j}{\partial t}. 
\ene
This is the first London equation, which 
reflects the dissipationless acceleration of superfluid. 

The second property that needs to be accounted for is the expulsion of
magnetic flux by a superconductor. The magnetic field 
is exponentially screened from the interior and hence follows a spatial
dependence 
\be \label{field-decay}
\nabla^{2}H = H/\lambda^{2}, 
\ene
where $\lambda$ is a screening
length called the London penetration depth. 
Together with the Maxwell equation $\nabla \times H = 4\pi j/c$, 
this implies the following condition between $H$ and $j$:
\be \label{london2}
H=- \left( \frac{4 \pi \lambda^{2}}{c}\right)
(\nabla \times j).
\ene
This is the second London equation and it describes the property of a
superconductor to exclude magnetic flux from its interior. The prefactors
in the two London equations are related through the Maxwell equations.
Taking the curl of both sides of \eqr{london1} and replacing $\nabla \times
E$ on the left side with 
$-\frac{1}{c}\frac{\partial H}{\partial t}$ and
substituting for 
$\nabla \times j$ on the R.H.S. with \eqr {london2} gives the expression
for the penetration depth as 
\be
\label{lambda}
\lambda = \sqrt{m^{*}c^{2}/4\pi n^{*}_{s}e^{*2}}
= \sqrt{mc^{2}/4\pi n_{s}e^{2}},
\ene
where the pair quantities with asterisks have been replaced with their
more common single-carrier counterparts ($2n^{*}_{s}=n_{s}$, $e^{*}=2e$,
and $m^{*}=2m$) in the last step\footnote{In subsequent expressions we retain
the asterisk marked quantities in our notation to maintain
generality. This allows, for example, the inclusion of 
band-structure effects on the effective
mass so that a more exact value of $m^{*}$ can be used rather than
simply taking the $m^{*}=2m$ corresponding to the free-electron model.}.
In the MKSA (SI) system, \eqr {lambda} becomes 
$\lambda = \sqrt{10m^{*}/4\pi n^{*}_{s}e^{*2}}$. 

A third relationship of importance in the discussion of the
pair-breaking current concerns the thermodynamic critical field \hcy . 
When flux is expelled, the free energy density is raised by the amount
$H^{2}/8\pi$. The critical flux expulsion energy 
(for the ideal case of a type I superconductor with a
non-demagnetizing geometry and dimensions large compared to the
penetration depth) corresponds to the condition
\be \label{fc}
f_{c} = f_{n} - f_{s} = H_{c}^{2}/8\pi
\ene
where the L.H.S. of the equation represents the condensation energy---the 
difference in free energy densities $f_{n} - f_{s}$ between the normal
and superconducting states. 

\subsubsection{Depairing current density in the London framework}

We now have all the ingredients we need to obtain the usual London 
estimate of the depairing current density. Taking \jd to represent the
condition when the kinetic energy density equals the condensation
energy, we have $ \frac{1}{2} n^{*}_{s}m^{*}v_{s}^{2} 
=\frac{1}{2}m^{*}j_{d}^{2}/
(n^{*}_{s}e^{*})^{2}= f = H_{c}^{2}/8\pi$. Substituting for 
$\lambda$ (Eq.~\ref{lambda}) then gives the required expression for the
depairing current density
\be
\label{london-jd}
j_{d} = \frac{cH_{c}}{4\pi\lambda}.
\ene
In practical MKSA
units, \eqr {london-jd} can be written as 
$j_{d}=(10^{7}/4\pi)(B_{c}/\lambda)$, where $j_{d}$ is
in A/m$^{2}$, $B_{c}$ is the thermodynamic critical flux 
density in Teslas and $\lambda$ is the penetration depth in meters.

Note that this derivation assumed that $n_{s}$ remains unchanged 
as $j$ approached \jdy . In reality \ns diminishes as the
superconductivity is destroyed and the normal phase boundary is
approached. Hence Eq.~\ref{london-jd} will be an overestimate for \jdy . 

\subsection{Ginzburg-Landau theory} 
\subsubsection{Basic framework}

There are situations where a system's quantum wavefunction cannot be solved for
by usual means because the Hamiltonian is unknown or not easily approximated. 
The GL formulation is a clever construction that allows 
useful information and conclusions to be extracted 
in such a situation where one cannot solve the problem quantum mechanically.
For describing macroscopic properties---such
as \jd that we are about to calculate---the GL theory is in fact 
more amenable than the microscopic theory. 

The idea is to introduce a complex phenomenogical
order parameter (pseudowavefunction) $\psi=|\psi|e^{i \varphi}$
to represent the superconducting state. $|\psi (r)|^{2}$ is
assumed to approximate the local density of superconducting charge carriers
$n_{s}(r)$. The free energy density $f_{s}$ of the superconducting state is
then expressed as a reasonable function of  $\psi (r)$ plus other energy terms.
A ``solution'' to  $\psi (r)$ is now obtained by 
the minimization of free energy rather than through quantum mechanics.
The unknown parameters of the theory are then solved in terms of measurable
physical quantities thereby providing contraints between the different
quantities of the superconducting state. 

Close to the phase boundary, \psay $^{2}$ is small and so $f_{s}$ can be
expanded keeping the lowest two orders of \psay $^{2}$. 
First let us consider the simplest situation where there are no
currents, gradients in \psay , or magnetic fields present. Then we have 
\begin{equation} \label{simpleGL}
f_{s}= f_{n} + \alpha |\psi|^{2} + \frac{\beta}{2} |\psi|^{4}, \end{equation}
where $\alpha$ and $\beta$ are temperature dependent coefficients whose
values are to be determined in terms of measurable parameters. 
The coefficients can be determined as follows. First of all for the solution
of   $|\psi|^{2}$ to be finite at the minimum free energy, $\beta$ must
be positive. Second, for the solution of $|\psi|^{2}$ to be  non-zero,
$\alpha$ must be negative. Since $|\psi|^{2}$ vanishes above \tcy , 
$\alpha$ must change its sign upon crossing \tcy . 
The minimum in $f_{s}$ occurs at
\be \label{GLsol1}
  |\psi|^{2}=-\alpha/\beta. 
\ene
Substituting this back in Eq.~\ref{simpleGL} and using the definition of
\hc (Eq.~\ref{fc}), Eq.~\ref{GLsol1} can be recast as
\be \label{hc-alpha}
f_{c} = \frac{H_{c}^{2}}{8\pi} = \frac{\alpha^{2}}{2\beta} 
\ene
giving one of the connections between $\alpha$ and $\beta$ and 
a measurable quantity (\hcy ). 
A second connection can be obtained by noting
that $n^{*}_{s}$ in Eq.~\ref{lambda} can be replaced by \psay $^{2}$, taking it's
equilibrium value from Eq.~\ref{GLsol1}
\be \label{lambdaGL}
\lambda^{2} = \frac{m^{*} c^{2}}{4\pi |\psi|^{2} e^{*2}} = 
\frac{-\beta}{\alpha}
\left(\frac{m^{*}c^{2}}{4\pi e^{*2}} \right).
\ene
Solving Eqs.~\ref{hc-alpha} and \ref{lambdaGL} simultaneously gives the 
GL coefficients:
\be \label{alpha-beta}
\alpha = -\frac{e^{*2}}{m^{*}c^{2}}H_{c}^{2} \lambda^{2}
\mbox{\hspace{2em} and 
\hspace{2em}} 
\beta=\frac{4\pi e^{*4}}{m^{*2}c^{4}}H_{c}^{2}\lambda^{4}
\ene

Next, the effect of fields and currents can be included in Eq.~\ref{simpleGL}
by adding terms corresponding to  the field energy density and 
kinetic energy of the current: 
\begin{eqnarray} \label{GLfree} 
-f_{c}\>= \alpha |\psi|^{2} + \frac{\beta}{2} |\psi|^{4} +
\frac{1}{2}|\psi|^2 m^{*}v_{s}^{2} + \frac{H^{2}}{8\pi} 
\nonumber \\
\>=\alpha |\psi|^{2} + \frac{\beta}{2} |\psi|^{4} 
+ \frac{1}{2m^{*}}\left|\left(\frac{\hbar}{i}\nabla - \frac{e^{*}}{c}A\right)
\psi\right|^{2}
+ \frac{H^{2}}{8\pi} \end{eqnarray}
where $\mbox{\boldmath $v$}_{s}$ is the superfluid velocity.
If the amplitude $|\psi|$ is constant and only its phase $\varphi$ varies
spatially, then $\mbox{\boldmath $v$}_{s}= (\frac{\hbar \nabla \varphi}{m^{*}}
-\frac{e^{*}\mbox{\boldmath $A$}}{cm^{*}})$. 
Also in the situation of interest to us---currents of pair-breaking
level---the gradient term is much larger than the $A$ term in the
operator for velocity. In this case 
$\mbox{\boldmath $v$}_{s} \approx \frac{\hbar \nabla \varphi}{m^{*}}$.

It should be noted that the above derivation assumed proximity
to \tc only for the purpose that \psa be small 
so that $f_{s}$ could be represented as a
limited power series expansion.  
In ``dirty'' superconductors---superconductors with a high impurity
scattering rate---the approximate validity of the GL expressions extends down
to $T \ll$ \tcy . In general, the expressions should be 
valid at all $T$ as long as \psa is small 
and the proper temperature dependent values of $\alpha$ and $\beta$ are
used, as expressed through the temperature dependencies of \hc and
$\lambda$. 
The treatment thus far assumes that charge carriers from only one band
contribute to the superconductivity, i.e., it is a SBGL (single-band
Ginzburg-Landau) theory. 

\subsubsection{Depairing current density in the SBGL framework}
We can now derive \jd by finding the value of $j$ above which
\psa vanishes. First we consider the case when $H=0$. 
We will assume $j$ to be uniform across the cross section. We will
justify this assumption later and look in detail at the conditions 
when the assumption is valid. Eq.~\ref{GLfree} then simplifies to 
\begin{equation} \label{simpleGLfree}
-f_{c} = 
\alpha |\psi|^{2} + \frac{\beta}{2} |\psi|^{4} +
\frac{1}{2}|\psi|^2 m^{*}v_{s}^{2}. \end{equation}
For zero $v_{s}$, we saw earlier  (Eq.~\ref{GLsol1}) that 
the equilibrium value of $|\psi|^{2}$ that minimizes the free
energy is $|\psi_{\infty}|^{2}=-\alpha/\beta$. For a finite $v_{s}$
minimization of Eq.~\ref{simpleGLfree} gives 
\begin{equation}
|\psi|^{2}= \frac{-\alpha}{\beta} \left( 1 - \frac{m^{*}v_{s}^{2}}{2
|\alpha|} \right) \end{equation}
with the corresponding supercurrent density 
\begin{equation} \label{nonmonoj}
j= e^{*}|\psi|^{2}
v_{s}
= \frac{-e^{*} \alpha}{\beta}
\left(1 - \frac{m^{*}v_{s}^{2}}{2 |\alpha|} \right) v_{s}. 
\end{equation}
The maximum possible value of this expression can now be identified with
$j_{d}$: 
\begin{equation} 
\label{SBGL-jdt1}
j_{d}(T)= \frac{-2e^{*}\alpha}{3\beta}
\left(\frac{2|\alpha|}{3m^{*}} \right)^{1/2} = 
\frac{c H_{c}(T)}{3\sqrt{6} \pi \lambda (T)}
\end{equation}
where the GL-theory parameters were replaced by their 
expressions in terms of the physical measurables \hc and $\lambda$ 
through Eqs.~\ref{alpha-beta}. 

The approximate temperature dependence of \jd can be obtained by inserting
the generic empirical temperature dependencies 
$H_{c}(T) \approx H_{c}(0)[1-(T/T_{c})^{2}]$ and 
$\lambda(T) \approx \lambda(0)/\sqrt{[1-(T/T_{c})^{4}]}$ giving
\begin{equation} \label{jdtfull} 
 j_{d}(T) \approx j_{d}(0) [1-(T/T_{c})^{2}]^{\frac{3}{2}} 
[1+(T/T_{c})^{2}]^{\frac{1}{2}} 
\end{equation}
where 
\begin{equation} \label{jdzero}
j_{d}(0)=cH_{c}(0)/[3\sqrt{6}\pi \lambda(0)] \end{equation}
is the zero-temperature depairing current density. In practical MKSA
units, \eqr {jdzero} can be written as 
$j_{d}(0)=10^{7} \times B_{c}(0)/[3\sqrt{6}\pi \lambda(0)]$, where $j_{d}$ is
in A/m$^{2}$, $B_{c}$ is in T and $\lambda$ is in m.

For the dirty case, instead of \eqr {jdtfull}, 
the temperature dependence of \eqr {SBGL-jdt1}
can be
better approximated as \cite{bardeen,romijn}
\begin{equation} \label{jdtfull-dirty} 
 j_{d}(T) \approx j_{d}(0) [1-(T/T_{c})^{2}]^{\frac{3}{2}}. 
\end{equation}

Usually $H_{c}$ can't be measured reliably, but 
from the relation 
\be \label{hc-hc2-lambda}
H_{c}=\sqrt{\frac{\Phi_{0} H_{c2}}{4\pi \lambda^{2}}}
\ene
\eqr {jdzero} can be recast as
\be \label{jdzeroGL}
j_{d}(0)=\sqrt{\frac{ c^{2}\Phi_{0} }{ 216\pi^{3} } }
\lb \frac{\sqrt{H_{c2}(0)}}{\lambda^{2}(0)} \rb ,
\ene
which in MKSA becomes 
$j_{d}(0)= 5.56 \times 10^{-3} \times \sqrt{B_{c2}(0)}/\lambda^{2}(0)$, 
where $j_{d}$ is in A/m$^{2}$, $B_{c2}$ is the upper critical flux 
density in Teslas and $\lambda$ is in meters.

Close to \tcy , Eq.~\ref{jdtfull} reduces to 
$ j_{d} (T \approx T_{c}) \approx 
4j_{d}(0)[1-T/T_{c}]^{\frac{3}{2}}$. This can be inverted to give 
the shift in transition temperature
$T_{c2}(j)$ at small currents, with the well-known $j^{2/3}$ proportionality:
\begin{equation} \label{tcjsmall}
1 - \frac{T_{c2}(j)}{T_{c}} \approx 
\left(\frac{1}{4}\right)^{\frac{2}{3}} 
\left[\frac{j}{j_{d}(0)}\right] ^{\frac{2}{3}}, \end{equation}
where $T_{c} \equiv T_{c2}(0)$. 
Note that if heat removal from the sample is ineffective, 
Joule heating will give an apparent shift $\Delta T_{c} \propto \rho
j^{2}$, which is the cube 
of the intrinsic $\sim j^{2/3}$ depairing shift near \tcy . Hence Joule
heating is easily distinguishable from a pair-breaking shift. 
(The  preceding discussion is based on Refs. 
\onlinecite{tinkhamtext} and \onlinecite{bardeen}.) 

\subsection{GL formulation for a two-band superconductor}
We now consider the two-band Ginzburg-Landau (2BGL) applicable to a 
system such as \mgb where two bands contribute to condensates. 
In this case, the condensation energy density can be expressed as
\cite{gurevich,askerzade,askerzade2} 
\begin{eqnarray} \label{2BGL-free}
-f_{c}= \left\{ \alpha_{1} |\psi_{1}|^{2} + \frac{\beta_{1}}{2} |\psi_{1}|^{4} 
+ \frac{1}{2m_{1}^{*}}\left|\left(\frac{\hbar}{i}\nabla - \frac{e^{*}}{c}A\right)
\psi_{1} \right|^{2} \right\} \nonumber \\
+ \left\{ \alpha_{2} |\psi_{2}|^{2} + \frac{\beta_{2}}{2} |\psi_{2}|^{4} 
+ \frac{1}{2m_{2}^{*}}\left|\left(\frac{\hbar}{i}\nabla - \frac{e^{*}}{c}A\right)
\psi_{2} \right|^{2} \right\} \nonumber\\
+ \left\{ \epsilon [\frac{}{}\psi_{1}^{*}\psi_{2} + c.c.] \right\} 
+ \left\{ \frac{H^{2}}{8\pi} \right\} 
\end{eqnarray}
where the first two braces correspond to the free energy contributions of
the condensates of each band and the third term corresponds to the interband 
interaction energy.

As before, for the case of no applied field and uniform current distribution,
this simplifies to
\begin{eqnarray} \label{2BGL-free2}
-f_{c}= \left\{ \alpha_{1} |\psi_{1}|^{2} + \frac{\beta_{1}}{2} |\psi_{1}|^{4} 
+ \alpha_{2} |\psi_{2}|^{2} + \frac{\beta_{2}}{2} |\psi_{2}|^{4}
\right\} \nonumber \\
+ \frac{1}{2}\left\{ \frac{}{} |\psi_{1}|^{2}m_{1}^{*}v^{*2}_{s1} +
|\psi_{2}|^{2}m_{2}^{*}v^{*2}_{s2} \right\}
+ \epsilon\left\{ \frac{}{}\psi_{1}^{*}\psi_{2} + c.c. \right\}.
\end{eqnarray}
The phases of the two condensates are locked together since 
at equilibrium the interband free energy is
minimized  when $\cos (\phi_{1} -\phi_{2})=1$ or $-1$ (for $\epsilon <0$
and $\epsilon >0$ respectively). 
Hence the superfluid momenta are equal, $m_{1}^{*}v^{*}_{1}=m_{2}^{*}v^{*}_{2}$,
and the superfluid velocities, $v^{*}_{1}$ and $v^{*}_{2}$, will be similar 
to the extent that the effective masses are similar. 

Because of the rather large number of parameters in Eq.~\ref{2BGL-free2}, 
it is not very meaningful to derive an expression for \jd for direct 
quantitative comparison with the experimental data. Rather we
will take the SBGL expression for \jd in Eq.~\ref{SBGL-jdt1} 
and insert the actual measured temperature dependencies
of \hc and $\lambda$. We expect those empirical temperature dependencies
to account for modifications due to the presence of two bands. As will
be seen below, the experiment confirms this contention. 
Note that for current flow
in the {\em ab} plane (which is our experimental situation) the stronger
planar $\pi$ band provides the main contribution to $\j_{s}$. In this
case the behaviour should qualitatively follow SBGL with the
appropriately modified parameters. 
 

\subsection{Depairing current from quasiparticle-energy shift}

In the previous derivations in the framework of phenomenological theories, no 
account was taken of the superconducting gap. 
As a final step in obtaining theoretical estimations of \jdy , let us
consider how the supercurrent is limited because of its effect on the 
superconducting gap. Unlike the GL
treatments, this one is particularly applicable to the $T \ll$ \tc
regime. In the microscopic theory, the superfluid density does not
decline continuously as $j$ and $v_{s}$ are increased. Rather, \ns
remains roughly constant until \vs reaches its depairing value
\cite{tinkhamtext,bardeen} 
\be \label{vsc}
v_{d} = \frac{\Delta}{\hbar k_{F}}=\frac{2\hbar}{\pi m^{*} \xi_{0}},
\ene
which corresponds to the point when the shift, $\hbar {\bm k_{F}.\bm v_{s}}$,
in the quasiparticle energies (QPE) equals the gap causing the gap to vanish
for those states. 
In Eq.~\ref{vsc}, $\xi_{0}$ is the intrinsic BCS coherence length.
The value of $\xi$ deduced from the upper critical field will be a lower
bound on $\xi_{0}$ since scattering reduces the effective $\xi$ (in the 
dirty-limit $\xi \approx \sqrt{\xi_{0}l}$, where $l$ is the mean-free path
\cite{tinkhamtext}).
$j$ remains closely proportional to \vs until $v_{d}$, then \ns drops 
precipitously 
so that the maximum current is only slightly (2\%) higher than the value when
\vsy =$v_{d}$. Thus to a good approximation $j_{d}\simeq  e^{*}n^{*}_{s}
v_{d}$. Combining this with \eqr {vsc} gives 
\be 
\label{nsvs-jd}
j_{d}\simeq  
\frac{m^{*} c^{2}\Delta}{4\pi e^{*}
\lambda^{2} \hbar k_{F}} 
\simeq 
\frac{c^{2}\hbar}{2\pi^{2} e^{*}\lambda^{2} \xi_{0}}.
\ene

The first R.H.S. involves the parameters such as $\Delta$ and $k_{F}$
whose absolute values are not known accurately but whose \tdepi are
well established, since $k_{F}$ is of course constant and 
the gaps have a temperature dependence that is well 
described by the standard BCS function 
\cite{silee-gap,canfield-phystoday}. 
Hence from the first R.H.S. we determine the \tdep of \jd as 
\be
\label{nsvs-jdt}
j_{d}(T) =j_{d}(0)
\lb \frac{\lambda^{2}(0)}{\Delta(0)}\rb 
\frac{\Delta(T)}{\lambda^{2}(T)}.
\ene

On the other hand, the second R.H.S. of Eq.~\ref{nsvs-jd} involves parameters 
such as $\xi_{0}$ whose absolute magnitude is better known (from
measurements of \hcu and the relation $H_{c2}=\Phi_{0}/2\pi \xi^{2}$).
Hence we use the second R.H.S. 
of \eqr {nsvs-jd} to estimate the absolute magnitude of \jdy$(0)$ as
\be \label{nsvs-jd0}
j_{d}(0) = \lb \frac{c^{2} \Phi_{0}}{8\pi^{5}} \right)^{\frac{1}{2}}
\frac{\sqrt{H_{c2}(0)}}{\lambda^{2}(0)},
\ene
which in MKSA becomes 
$j_{d}(0)= 9.19 \times 10^{-3} \times \sqrt{B_{c2}(0)}/\lambda^{2}(0)$.

\subsection{Microscopic calculation} 
Various authors have calculated \jdt from a microscopic basis
\cite{bardeen,maki,ovchinnikov}. For
arbitrary temperatures and mean free paths, 
one must use the Gorkov equations as the starting point. Kupriyanov and
Lukichev \cite{KL} have derived \jdt from the Eilenberger equations,
which are a
simplified version of the Gorkov equations. This derivation is beyond
the scope of the present review, but a nice shortened version can be
found in reference \onlinecite{romijn}. The microscopic calculation
confirms the overall temperature dependence predicted by GL and
the two normalized curves differ only slightly from each other 
(e.g., see Fig.~4 of reference \onlinecite{romijn}). 

\subsection{Comparison between different theoretical estimates}
We will now compare the different theoretical estimates of
\jdy$(0)$ obtained by the different approaches and calculate their relative
ratios from Eqs.~\ref{london-jd}, \ref{jdzero}, \ref{jdzeroGL}, and 
\ref{nsvs-jd0}. The London estimate, as
discussed earlier, should be an overestimate. Its value is 1.84 times
higher than the GL estimate. Similarly, the estimate from QPE shift
turns out to be 1.65 times higher than GL. Since the QPE shift
calculation is based on a simple single isotropic gap, the actual
\jdy$(0)$ will be different and somewhat lower. For the very dirty 
($l \ll \xi_{0}$) and very clean ($l \gg \xi_{0}$) limits, the actual \jdy$(0)$
is expected to be 0.67 and 0.82 times the London estimate respectively
\cite{bardeen}. 
Hence these estimates
should be correct at most to a factor-of-two accuracy. Errors in the
values of the parameters will further add to the inaccuracy of the
calculated value. So when comparing with the experiment, an agreement to
within half an order of magnitude is about the best that can be
expected. 

\section{Experimental methods}
\subsection{Samples} 
The samples are 400nm-thick {\em c}-axis oriented films of MgB$_{2}$ on
sapphire. In this paper we show data on four bridges, labelled S, M, N, and L 
with lateral dimensions 2.8 x 33, 3.0 x 61, 3.0 x 60, and 9.7
x 172 $\mu$m$^{2}$ respectively. The lateral
dimensions are uncertain by $\pm 0.7 \mu$m and the mean
thickness by $\pm 50$ nm.
TEM (transmission electron microscopy) shows variations in thickness
of about $\delta t/t\sim 10$\%. The films were
fabricated using a two-step method whose details are described 
elsewhere\cite{sampleprep,sampleprep2}. An amorphous boron film was 
deposited on a (1\={1}02) Al$_{2}$O$_{3}$ substrate at room temperature by 
pulsed-laser ablation. The boron film
was then put into a Nb tube with high-purity Mg metal (99.9\%)
and the Nb tube was then sealed using an arc furnace in an
argon atmosphere. Finally, the heat treatment was carried out at
900$^{\circ}$ C for 30 min. in an evacuated quartz ampoule 
sealed under high vacuum. X-ray diffraction indicates
a highly c-axis-oriented crystal structure normal to
the substrate with undetectable ($<$ 0.1\%)
 impurity phases. Magnetization $M(T)$
curves have a $\lambda$-limited transition width of 1.5K (\tcy -spread
$<$ 0.2K). However, the aforementioned variations in thickness 
produce a broadening of $R(T)$ with increasing $j$.
The films were photolithographically patterned down to narrow bridges
and in the case of one sample (N) the lead areas were further delineated by
mechanical scribing. 

The samples have a 
normal-state
resistivity at room temperature $\rho_{n}(300 \mbox{K}) \approx 14
\mu\Omega$-cm, which is 
about 7 times
that found in clean single crystals \cite{ott}. The enhancement appears to 
to be due to microscopic disorder (scattering) and not just some
extra series resistance at grain boundaries, since
\hcu is also enhanced by about the same factor 
over its values in single crystals \cite{ott,mgbflow}. Also measurements of
$\lambda$ on these films show no evidence of grain-boundary
weak-linked behaviour, which would be manifested as a non-linearity in
response and as an increase in the apparent absolute magnitude of
$\lambda$, which were not observed \cite{kim}.

\subsection{Cryogenics}
Two sets of apparatus were used for the measurements. Samples S, M, and L were
measured in an Oxford Instruments$^{(R)}$ 
liquid-helium based vapour-flow cryostat
with a 16 Tesla superconducting magnet. Sample N was measured in a
Cryomech$^{(R)}$ pulsed-tube closed-cycle refrigerator and a water-cooled
copper magnet. In the latter case, the ``zero field'' could be made very
small ($<0.7$ G) by shielding the sample region using mu-metal.
Lakeshore$^{(R)}$ diode and cernox sensors were employed for the
thermometry. 

\subsection{Electrical measurements}
\begin{figure}[h] 
 \begin{center}
	\leavevmode
	\epsfxsize=0.8\hsize
	\epsfbox{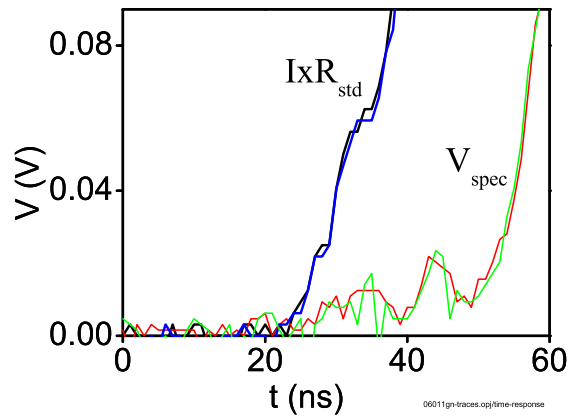}
 \end{center}
 \vspace{-2em}
\caption{\label{time-response}{Rising edges of voltage and 
current (voltage across R$_{\mbox{std}}$) across an \mgb bridge (sample
N). The sample is in
the normal state at room temperature. The graph contains two sets of
measurements of $I(t)$ and $V(t)$ showing overlap and temporal
reproducibility to within about 1 ns.}}
\end{figure}

Low-current resistance measurements ($I < 50 \mu$A) were made by the 
standard 
four-probe DC technique. Higher current measurements were carried out
using a pulsed technique with a four-probe configuration.
A pulsed signal source (capable of both constant-current and constant-voltage
modes) drives the signal through the sample and a series standard
resistor R$_{\mbox{std}}$.  The signals across the sample and R$_{\mbox{std}}$
are then detected with a digital-storage oscilloscope. The temporal
reproducibility of the system is \aby 1 ns as can be evidenced from \figr
{time-response}, which shows two sets of measurements of $I(t)$ and $V(t)$.
The relative delay between rising edges of $I(t)$ and $V(t)$ corresponds
approximately to the extra length of wires divided by the speed of light. 

\begin{figure}[h] 
 \begin{center}
	\leavevmode
	\epsfxsize=0.8\hsize
	\epsfbox{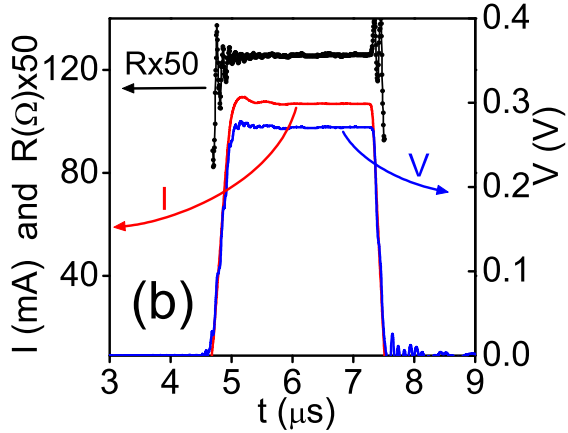}
 \end{center}
 \vspace{-2em}
\caption{\label{pulses}{Pulse waveforms at high dissipation levels 
($j = 9.7$ MA/\cmsy,
$E = 83$ V/cm, and  $p = jE = 803$ MW/\cmc on the plateaus). The
measurement was done on an \mgb bridge (sample S) just above \tc (at 42K).
 The resistance rises to (90\% of) its final value in about 50
ns from the (10\%) onset of $I$.}}
\end{figure}

Pulse durations range 0.1--4 $\mu$s with a duty cycle of about 1 ppm. 
About 100 pulses are averaged to reduce the noise. 
A typical pair of current and voltage
pulses is shown in Fig.~\ref{pulses}. 
The values of $I$ and $V$ used
in subsequent analyses correspond to the flat plateaus of the
pulse waveforms. As these are time independent, the measurement
corresponds effectively to a DC measurement albeit over a shorter-than-usual
duration. The computed $R(t)=V(t)/I(t)$, shown in
Fig.~\ref{pulses},
 is seen to have a 50 ns rise time. The total duration of the pulse is
not relevant but the time at which the voltage or current is measured 
since it turns on. Since the signals saturate to well defined values 
by \aby 100 ns, this is the effective dissipation time $\tau$ that is
relevant to the thermal calculations below. 
Some additional information on the electrical-measurement setup
can be found in a previous review article
\cite{mplb} and other recent papers \cite{metal,unstable}. 


\begin{figure*}
 \begin{center}
	\leavevmode
	\epsfxsize=0.75\hsize
	\epsfbox{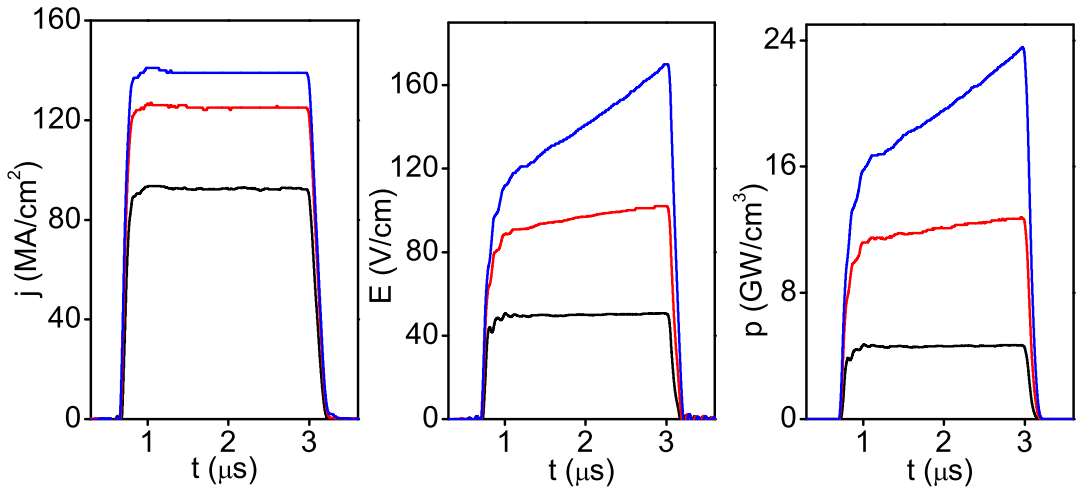}
 \end{center}
 \vspace{-2.5em}
\caption{\label{sloped-pulses}{Pulse waveforms showing a time-dependent
temperature rise at extreme dissipation levels ($p > 10^{10}$ W/\cmcy ).
The measurement was made on a 100 nm thick \ybco film
on LaAlO$_{3}$ \cite{metal,unstable} at $T$=50 K and $B$=10 T. The left,
middle, and right panels show the current density, electric field, and
power-dissipation density respectively.}}
\end{figure*}
\subsection{Heat conduction during pulsed measurements} 
We now look in more detail at 
the thermal processes involved in the removal of heat from the sample
\cite{nahum,gupta}. 
Heat generated in the film diffuses toward the interface with the
substrate essentially instantaneously.  On the time scale of nanoseconds,
phonons transfer heat across the interface between the film and
substrate. Heat then diffuses within the substrate in a matter of
microseconds and finally into the heat sink in milliseconds. Thus for pulses
of microsecond or shorter duration, the heat doesn't leave the
substrate and for low duty cycles the thermometer will not register a
temperature rise. 
The sample temperature rise is then composed of a temperature gradient
within the film, an abrupt temperature difference across the boundary,
and a gradient within the substrate. Each can be expressed as an
additive component of the thermal resistance \rthy , which we will
define as the temperature rise $\Delta T$ per unit power density $p$. 

 The temperature
variation within the film given by 
$\Delta T_{1}(z) \simeq (p/\kappa_{f})(tz-z^{2}/2) $
where $(t-z)$ is the distance of a point within the film from the
interface, $t$ is the film 
thickness and $\kappa_{f}$ is the thermal conductivity
of the film. Expressed as a thermal resistance this becomes 
\begin{equation} \label{rth1}
R_{th1}(z) \simeq \frac{1}{\kappa_{f}}\lb tz-\frac{z^{2}}{2}\rb. 
\end{equation}
For \mgby , $\kappa_{f}=0.1$ and 0.4 W/K.cm at 10 K and 40 K respectively
\cite{putti,muranaka}. This yields an average \rth in the midplane of the film
(i.e., at $z=t/2$) of 6 nK.\cmcy/W at 10K and 1.5 nK.\cmcy/W near \tcy . 

 Phonon mismatch at
the interface between film and substrate produces an abrupt temperature
drop $ \Delta T_{2}=pt R_{bd}$,
where $R_{bd}$ 
is the thermal boundary resistance at the film-substrate interface. 
This can be expressed as a second component to
\rthy :
\be \label{rth2} 
R_{th2}=t R_{bd}.
\ene
Most metallic films on sapphire exhibit the approximate 
empirical rule $R_{bd}T^{3}\approx 20$ \cmsy.K$^{4}$/W \cite{swartz}. Thus
$R_{th2} \sim 12$ nK.\cmcy/W at 40 K (near \tcy) and 
$\sim$800 nK.\cmcy/W at 10 K. (Ceramic on ceramic boundaries, such a
\ybco on LaAlO$_{3}$ seem to fair better in this respect.) 

 Finally as the heat pulse propagates through the substrate, there is a 
third component to \rthy , which for the long rectangular strip geometry is
 given by\cite{gupta}:
\begin{equation} \label{rth3}
R_{th3} = \frac{2.26t\tau
w}{{2(D_{s}\tau)^{1/2}[4(D_{s}\tau)^{1/2}+w]c_{s}}}, \end{equation}
where $D_{s}=\kappa_{s}/c_{s}$ is the diffusion constant, 
$\kappa_{s}$ and $c_{p}$ are the thermal conductivity and specific heat
of the substrate material, $w$ is the width of the bridge, and $\tau$ is
the duration for which the power is applied. For sapphire, $c_{s} \sim
3.2$ J/K.\cmc \cite{perkin-sichina} and $\kappa 
\approx 10$ and 100 W/cm.K at 10
K and 40 K respectively \cite{saphikon}. Thus $D_{s}=\kappa_{s}/c_{s} 
\sim 3$ and 30 \cmsy/s at 10 and 40 K respectively. This gives a thermal
diffusion lengths $\sqrt{D\tau}$ within the substrate of 6--17 $\mu$m for
$T$=10--40 K, for $\tau=100$ ns. Since our bridge
widths (\aby 2--10 $\mu$m) are small or comparable to these diffusion lengths,
the denominator \ab $8D \tau c_{s} \propto
\kappa_{s}$ is mainly dependent on the conductivity of the substrate and is
relatively independent of its specific heat. From Eq.~\ref{rth3} the last
component of the thermal resistance then has an estimated magnitude of 
 $R_{th3} \sim$ 0.8--0.1 nK.\cmcy/W in the $T$ range 10--40 K respectively 
for a 3 $\mu$m wide bridge. 

From the above estimated components, we can obtain the total thermal
resistance 
\be \label{rth}
R_{th} = R_{th1} + R_{th2} +R_{th3}
\ene
as \rth $=1.5 + 10 + 0.1 \approx 12$ nK.\cmcy/W near \tc and 0.8
$\mu$K.\cmcy/W at 10 K, where the 
dominating term comes from the boundary resistance. 


\subsection{Adiabatic heating}
The worst-case scenario for sample heating is when the timescales and
conditions are such that none of the generated heat escapes. In this
case the energy density $p\tau$
dumped by the pulse will go entirely into raising the internal energy
$U$. The temperature rise can then be expressed in terms of 
the film's specific heat as
\be \label{rth-adia-integral}
p\tau = \delta U = \int_{T_{0}}^{T'} c(T) dT \approx c\Delta T,
\ene
where the last approximation applies for small temperature shifts, $c\equiv
c(T_{0})$ is the specific heat at the nominal bath temperature  $T_{0}$,  and 
 $T'$ is the final raised temperature, i.e.,  $\Delta T =T'-T_{0}$. 
The effective thermal resistance in this case is 
\be
\label{rth-adia}
R_{thA} = \tau/c(T_{0}).
\ene
Close to \tcy, $c(40\mbox{K}) \approx 0.044$  J/K.\cmc  in \mgb
\cite{mgb2specheat}.  
For times between 100ns and 1$\mu$s after current turn on,  
the corresponding adiabatic \rth then has values in the range 
\aby 2--20 $\mu$K.\cmcy/W. Note that the $R_{thA}$ values are about three
orders of magnitude higher than when there is heat conduction (\eqr
{rth}) but place an absolute upper limit on the actual \rthy. Because 
 $R_{thA}$ is directly proportional to $\tau$, the amount of heating
can be made almost arbitrarily small by reducing the duration of
the signal (for very short durations there will be addional
corrections when the electrons can no longer equilibrate with the
phonons). The above estimates were for $T$\aby\tcy ,
 at 5K the corresponding $R_{thA}$ values will be one or two orders of
magnitude higher. 

\subsection{Heat generation at contacts} 
As seen above, pulsed measurements greatly reduce the effective thermal
resistance and accompanying temperature rise in the sample. In addition
they can essentially eliminate problems associated with heat produced at
lead contacts. This is
because the contacts are typically located several mm away from the
active part of the sample (the bridge) involved in the four-probe
measurement. This distance is typically large compared to 
the thermal diffusion distance $\sqrt{D\tau}$. The
specific heat at 10 K and 40 K is 1,400 and 50,000 J/K.m$^{3}$ 
respectively \cite{mgb2specheat,kremer-ahn}. Together with the values of 10 and 40
W/m.K for thermal conductivity \cite{putti,muranaka} 
gives diffusion constants of 0.007
and 0.0008 m$^{2}$/s at the two temperatures. Thus for dissipation
durations 0.1--1 $\mu$s and temperatures 10--40 K, we get diffusion distances
in the 9--84 $\mu$m range, which are about a hundred times shorter than
the distance to the bridge. So heat generated at the contacts does not 
interfere with the measurement.

\subsection{Pulse waveform distortion due to temperature rise}
The third component of the thermal resistance $R_{th3}$ arising from the
flow of heat into the substrate is time dependent (\eqr {rth3}). As a
result intense heating can cause the resistance to rise
with time and hence the voltage pulse becomes sloped \cite{mplb}. 
\figr {sloped-pulses} shows an example of this situation, where the 
measurements were done on \ybco films. Even
in the case of a system with a relatively $T$ independent \rrhon such as
the \mgb films, in the superconducting state the flux-flow resistivity
is $T$ dependent, especially in the regime near \tcy . 

For very short pulses and/or low enough power densities, the
temperature rise in the substrate $\Delta T=pR_{th3}$ becomes 
negligible and the main bottle neck for heat conduction becomes the
thermal boundary resistance. If these conditions are satisfied the voltage
pulse should appear flat. This was verified in all of our
measurements and data was only taken when the pulses were undistorted as
in \figr {pulses}. Note that the power-dissipation densities for the
pulses in \figr{sloped-pulses} are much higher than the levels 
used in the present work on \mgby . 

\subsection{Influence of thermal environment}
\begin{figure}
 \begin{center}
	\leavevmode
	\epsfxsize=0.8\hsize
	\epsfbox{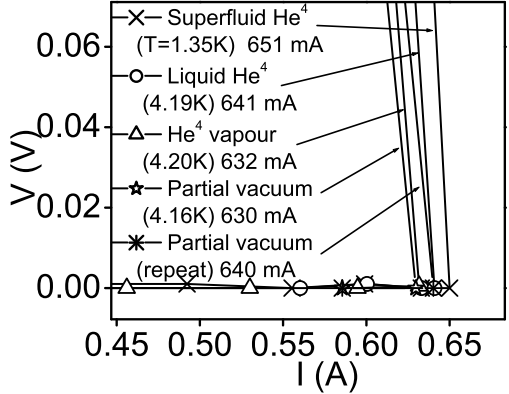} 
 \end{center}
 \vspace{-3em}
\caption{\label{thermal-env}{$IV$ 
curves for sample L 
at the lowest temperatures with the exposed surface of the film in
contact with different thermal environments as indicated.  The 
influence of thermal environment on Joule heating is seen to be minimal
as there is no major systematic change in the threshold current at which
the jump in voltage occurs.}}
\end{figure}

In measurements involving a continuous DC signal or long time scales, 
 substantial overall
heating can be alleviated by immersing the sample directly in a liquid
cryogen. This is especially noticeable when the cryogen is in a
superfluid state \cite{stoll}. 
In our measurements the timescales are such that even the
substrate does not experience a sigificant rise in $T$. In this case we
would expect that additional heat removal from the exposed surface of
the film to a cryogen to not make a big difference. This is indeed the
case as can be seen in \figr {thermal-env} where IV curves measured in different
thermal environments show a jump to the normal state at a 
value of $I_{d}$ that is roughly independent of the thermal environment. 

\subsection{Electron-phonon disequilibrium}
Beside macroscopic heating of the sample, other exotic situations can
arise where the electronic temperature is significantly raised with
respect to the phonons, which remain in equilibrium with the bath
\cite{LO,mus,klein,sam,doett,xiao,bezshkl,bez,volotskaya,gurevich-mints,
unstable,hote,eprelax}. The characteristic time for the relaxation of
energy between the electrons and phonons is typically under $10^{-8}$ s.
 This effect happens mainly when there is vigorous 
vortex motion and is not a main concern in the present work. 

\subsection{Temperature rise and thermal runaway}

Except in the case when the sample has perfect electrical conductivity, in the
dissipative state the measurement is necessarily accompanied by some
rise in sample temperature. This leads to the risk of thermal runway,
where a small temperature rise leads to increased dissipation, leading
to a further temperature rise and so on. The sample can
end up at a temperature much higher than the bath temperature and possibly in
the normal state. 
If this scenario is fulfilled, the quantity actually measured will be 
the conventional \jcy , demarkating the onset of dissipation, rather
than the true \jdy . The match needed to light a thermal runaway is
flux-flow dissipation whose power density is given by $p_{f} \alt 
j^{2} \rho_{n}B/H_{c2}$. 
A small localized hot spot will not sustain a runaway since the
current can simply avoid the hot spot and flow through 
more conductive regions. Also heat produced at one point does not remain
localized on the timescale of the measurement but spreads
uniformly across the width of the film because the thermal diffusion 
length, calculated earlier, is larger than the bridge width. Thus the
the relevant parameter is the average dissipation across the sample volume.

At 10 K, this $p_{f}$ is about 1.2 W/\cmc taking an average self field
of 100 G (all measurements here were made in zero applied magnetic field
with the sample chamber shielded by mu-metal in some cases). From the
\rth calculated earlier, this results in $\Delta T \approx 1.1$ K, which
increases the sample resistance only marginally (\rrhon of \mgb is
fairly flat; the rise occurs mainly because of a drop in \hcu with $T$). 
This will lead to an insignificant rise in  $p_{f}$ even for a constant
current. If the measurement is made with a constant-voltage source,  $p_{f}$
will actually drop with $R$ (since then $p_{f} \propto V^{2}/R$ 
instead of $p_{f} \propto I^{2}R$). Our apparatus, when set to constant-voltage
mode, has a source impedence as low as 0.3 Ohms 
which is about an order of magnitude below $R_{n}$. Thus the power
dissipated doesn't rise rapidly with $T$. In the meanwhile the thermal
properties that aid heat removal 
(conductivities and specific heat) become rapidly more effective 
as $T$ increases, working strongly to prevent a runaway. Thus it is
possible to make a quantitative evaluation of sample heating and the 
risk of thermal runaway. 

A more detailed discussion on
the subject of hot spots and thermal propagation in superconducting
microbridges can be found in the review article by Gurevich and Mints
\cite{gurevich-mints} and references therein. 

\subsection{Depinning current versus depairing current} 
As long as gross sample heating or thermal runaway can be eliminated as
discussed above, there is no confusion between the conventional
(depinning) \jc and \jdy. When \jd is
exceeded, the resistance reaches the normal-state value $R_{n}$. When
\jc is exceeded, the resistance approaches (if pinning is well
overcome) the free-flux-flow value $R_{f} \sim R_{n}B/H_{c2}$. The
latter is lower than   $R_{n}$ by three orders of magnitude. Hence the
critical current measured is not just the conventional depinning
threshold but a good estimate of \jdy . 

\subsection{Uniformity of current flow}
A superconducting wire with dimensions small compared
to $\xi$ and $\lambda$, automatically has a uniform distribution of $j$ and 
$|\Delta|$ across the cross section. Close to \tcy , $\xi$ and $\lambda$
diverge, so this condition applies to the \tcu shifts measured from
resistive transitions at fixed currents (\eqr {tcjsmall}). 

At lower temperatures and for sample cross sections that are not small,
the question of current-flow uniformity needs to be further examined.
In a type-I superconductor in zero applied field, the self flux is completely
expelled and the current flows without resistance 
along the periphery of the cross section up to some threshold value. 
Resistance first appears when the self field at the surface
exceeds \hc (Silsbee's rule). Beyond this the sample enters an
intermediate state where normal regions coexist with superconducting
ones and current flows through both so that the macroscopic
resistance is a fraction of the normal-state value. In this intermediate
state, the current is distributed more homogeneously over the sample volume. 

\begin{figure}
 \begin{center}
	\leavevmode
	\epsfxsize=0.7\hsize
	\epsfbox{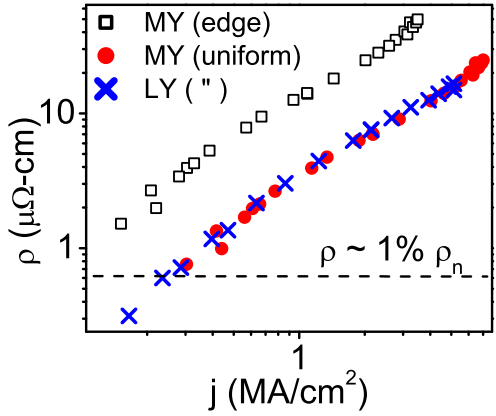}
 \end{center}
 \vspace{-2em}
\caption{\label{homo-flow}{Non-linear mixed-state response of two 90-nm
thick \ybco microbridges. Samples MY and LY are 8 $\mu$m and 16 $\mu$m
wide respectively. The lower two 
sets of symbols, showing \rrho and $j$ calculated
assuming uniform current flow, are in excellent agreement. 
The open squares at the top, showing \rrho and $j$ (for MY
relative to LY) based on an assumption of non-uniform current 
flow along edges, are in conspicuous disagreement. The dashed line corresponds
to 1\% of the normal-state resistivity.
The measurements were made at $T=50$ K and $B=13.8$T (\ab 20\% of \hcuy); 
 $\lambda \sim 200$ nm.}}
\end{figure}

In thin strips or bridges of type-II superconductors this process goes a
step further. First, instead of \hc you have a much lower \hcl that defines the
threshold of flux entry (possibly modified by surface barriers except
for thin film bridges). Likharev \cite{likharev} has shown that 
the minimum sample width for the nucleation of vortices is 
$4.4\xi(T)$. Second,
because of the highly aspected geometry and consequent large
demagnetization, the effective threshold is practically negligible. Thus
there are always flux filled regions within bridges wider than the
Likharev threshold. Also for the
type-II case, vortices are present that necessarily move since we are
concerned with current densities well beyond the depinning \jcy . 
Hence the current flow will become homogeneous
under these conditions. In general, as the
current grows beyond \jc and the system becomes highly resistive, the 
current flow becomes macroscopically uniform, as in a normal conductor, due
to the principle of minimum entropy production. This has been
investigated by us in our previous work on
\ybco bridges of different widths \cite{metal}. \figr {homo-flow} shows
$\rho(j)$ curves on two \ybco bridges measured in the highly non-linear
mixed-state regime. Note that the resistivity changes by almost two
order of magnitude and extends below 1\% of \rrhony . The values of 
\rrho and $j$ calculated based on uniform current flow (i.e., dividing
$I$ by the nominal cross section to obtain $j$, etc.) are in excellent
agreement. On the other hand if one assumes that the current flows
non-uniformly along the edges, the nominal cross section will
overestimate the actual area that contains the current. In this case,
the relative error between samples MY and LY will be a factor of two 
in \rrho and a factor of half in $j$ (the sample widths differ by a
factor of half). The edge flow assumption is checked by 
comparing the squares to the crosses. Clearly the assumption fails. 
We consider this to be definitive proof that the current does not just
flow along the edges but occupies the whole sample cross section
uniformly once the transport becomes dissipative. Other authors 
\cite{rusanov,geers} have also reached the same conclusion that 
vortex motion homogenizes current flow. 

It will be seen in the results below that the values of \jdy $(0)$ in \mgb
obtained from resistive-transition shifts near \tc agree with the
values obtained at low temperatures from $IV$ jumps. Also the values
obtained for samples with different cross sections are all in mutual
agreement. If the current flow did not fill the cross section, the
macroscopically calculated value of \jd would be higher for a narrower
sample. All of these observations confirm that current flow becomes
homogeneous under dissipative conditions.

\section{Results and analysis} 
\subsection{Resistive transitions at fixed currents}
\begin{figure}[h] 
 \begin{center}
	\leavevmode
	\epsfxsize=0.95\hsize
	\epsfbox{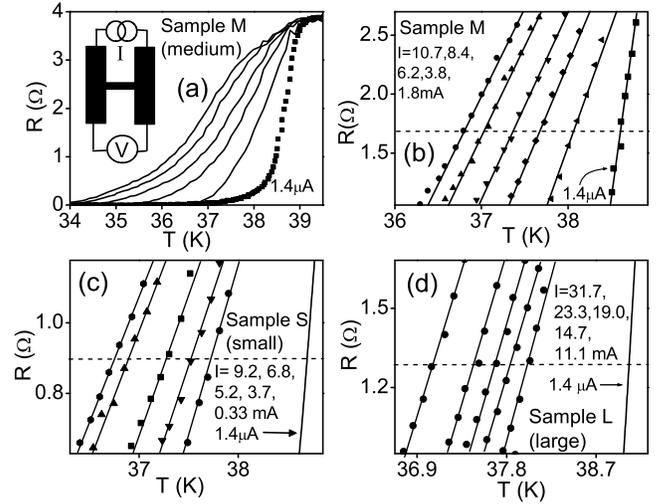}
 \end{center}
 \vspace{-1em}
\caption{\label{rtcurves}{Resistive transitions of \mgb bridges at
different currents (values correspond to curves from left to right.). 
Panels (a) and (b) show two windows of the same data. The inset in (a) shows
the sample geometry and configuration of 
leads. Panels (b), (c), and (d) show the
central main portions of the transitions for three different sized 
samples. The rightmost curves at I=1.4 $\mu$A were measured with a
continuous DC current; the rest used 
pulsed signals. The dashed lines represents $R=R_{n}/2$ for each sample.
Reprinted with permission from M. N. Kunchur, S.-I. Lee, and W. N. Kang 
Phys Rev. B {\bf 68}, 064516 (2003). (c) (2003) 
The American Physical Society.}}
\end{figure}
Fig.~\ref{rtcurves}(a) shows the resistive transitions at different electric
currents $I$
for sample M. The inset shows the sample
geometry. The horizontal sections of the current leads add a small
($\sim 15$ \%) series resistance 
to the actual resistance of the bridge. Because $j$ in these
wide regions is negligible, this resistance freezes out at the nominal
unshifted \tcy ,  making the onset seem to not shift. Similarly,  
the lower foot of the transition will have a flux-motion 
contribution $R_{f} \sim R_{n}B/H_{c2} < 5$\% $R_{n}$
from the self field. 
The central two-thirds portion of the transitions (magnified in
panel (b)) circumvents these errors, displaying relatively parallel
shifts due to pair breaking. 

Variations in film thickness cause the
transitions to broaden slightly with increasing $j$ despite phase
purity. For a simple model with series thickness variations $\delta t$, 
the functional shape of the current-dependent broadened transition 
is given by $R(j,T,\delta t) = R_{n}\{\log(j/4j_{d}(0)) - 1.5 \log (1-
T/T_{c})\}/\log (1+ \delta t/t)$, where $\delta t$=thickness variation.
At the $R_{n}/2$ criterion (shown by the dashed line) the actual  \tc
shifts correspond exactly to shifts for a sample with the same mean thickness
$t$ but with $\delta t=0$. 

Panels (c) and (d) show similar sets of curves for two more samples.
It should be noted that in addition to transition broadening due to
thickness variations, the $R(T)$ transition may have some
intrinsic width as a function of $j$. An analogous situation arises for
$R(T)$ in a magnetic field, where there is an intrinsic
broadening as $B$ is increased \cite{tinkhamrt}. Unfortunately there is
no theoretical work on the $R(T)$ transition shape at high $j$.
Nevertheless, we expect the midpoint criterion for the shifted \tc to
provide a factor-of-two estimate of \jdy .
\begin{figure}[h] 
 \begin{center}
	\leavevmode
	\epsfxsize=0.999\hsize
	\epsfbox{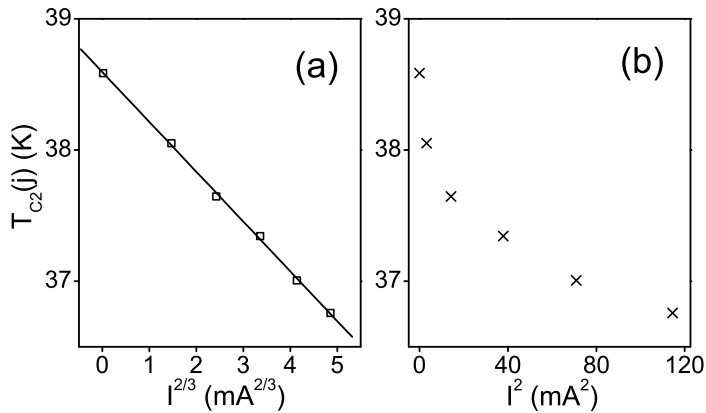}
 \end{center}
 \vspace{-1em}
\caption{\label{laws}{ 
Shifted transition temperatures 
at different currents. The two panels compare the
same \tcuy $(j)$ data versus $I^{2/3}$ and  $I^{2}$, showing adherance to
the  $I^{2/3}$ law for pair breaking rather than the
$I^{2}$ law for Joule heating. The linear fit (solid line) to the  
$I^{2/3}$ plot gives $I_{d}(0)= 257$ mA (see Eq.~\ref{tcjsmall}).}}
\end{figure}
Fig.~\ref{laws}(a) shows the midpoint \tcuy 's 
and their corresponding currents (ranging 
from $10^{-6}$ to $10^{-2}$ A) plotted as $I^{2/3}$ (expected for
pair-breaking) and as $I^{2}$ (expected for Joule heating).
 The shifts are closely proportional to $I^{2/3}$ 
rather than to $I^{2}$, showing that heating is not appreciable 
(the plots for the other samples look similar).  In fact from our
earlier calculation of the thermal resistance, we can estimate the expected
temperature rise. For the average $j \approx 0.4$ MA/\cmsy , and 
transition-midpoint
$\rho \approx 3 \mu\Omega$-cm, we get $p \approx 0.5$ MW/\cmcy . This
heats up the sample by $\Delta T = pR_{th} \approx 6$ mK. This is much 
smaller than the size of the symbols on the plot and two orders of magnitude
smaller than the observed shifts. Therefore heating is definitely negligible
for the data shown in Fig.~\ref{laws}.

The slope $d I^{2/3}/dT_{c}(j)$ together with Eq.~\ref{tcjsmall} gives
a zero-temperature depairing current value of 257 mA 
(if the \tc criterion is taken at 30\% and 70\% of $R_{n}$, the
corresponding \id values are 196 mA and 299 mA respectively). 
Dividing this by the cross-sectional area gives  \jdy $(0)$.
For samples S, M, N, and L the respective values of \jdy $(0)$ are 
$2.2 \times 10^{7}$, $2.1 \times 10^{7}$, $2.0 \times 10^{7}$, and
$1.8 \times 10^{7}$ A/\cmsy . 
The four values are consistent within the uncertainties in 
the sample dimensions, implying a cross-sectionally uniform current density.
As discussed earlier, 
this is especially expected close to the \tcuy$(j)$ boundary
where $\lambda$ and $\xi$ diverge and 
the superconductor becomes highly  dissipative due to flux
motion and fluctuations.

\subsection{Current-voltage curves at fixed temperatures}
\begin{figure}[h] 
 \begin{center}
	\leavevmode
	\epsfxsize=0.8\hsize
	\epsfbox{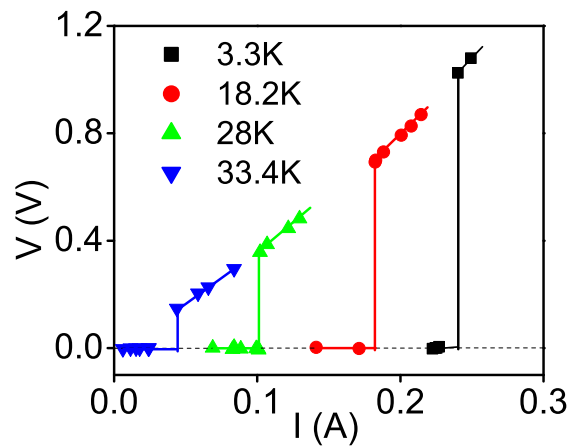}
 \end{center}
 \vspace{-1em}
\caption{{$IV$ 
curves for sample N at fixed temperatures (listed for curves going from
right to left). Lines are drawn to guide the eye. Curves at intermediate
temperatures were omitted for clarity. Beyond some threshold, 
the voltage jumps abruptly to the normal
state. The slanted portions have the slope $V/I = R_{n}$ and their
intercepts are zero.}}
\label{ivcurves}
\end{figure}
Fig.~\ref{ivcurves} shows current-voltage ($IV$) characteristics at various
fixed temperatures for sample N (results for samples S, M, and L are similar).
As $I$ is increased, $V$ remains close to zero until some
threshold value. Above this it switches abruptly to the Ohmic behaviour 
$V=I R_{n}$. This threshold is a lower bound for \idy ; however, 
as per our earlier discussion regarding thermal runaway, the temperature
error is of the order of 1 K at low temperatures and negligible at
higher temperatures. So the measured threshold can be associated with 
 \idy $(T)$ at the nominal bath temperature.

For $T \agt 35$ K the transition is gradual 
 whereas at the lower temperatures it is rather abrupt 
 \cite{mgbpair}. 
This is partly because of film thickness variation as discussed earlier and 
partly because a type II superconducting phase transition changes from
second order to first order at lower temperatures in the presence of a
current \cite{bardeen}. 

The measurements on sample N were done with a
signal source with a high source impedance thereby eliminating 
the ``s'' shape seen earlier in reference \cite{mgbpair}, where 
the external circuit had a source impedance comparable to that of the
sample. 

\subsection{Zero-temperature values of \jdy } 
From such $IV$ characteristics measured at the lowest temperature 
the current required to drive the sample normal
provides a direct estimate of \jdy$(T\approx 0)$. 
These values of \jdy$(0)$ for the four samples are 
shown in Table~1 (top row of numbers).  
\begin{table}[h]
\begin{tabular}{||l|r|r|r|r||}
\hline \hline
&\multicolumn{4}{|c||}{Sample}\\
\cline{2-5}
Method:&S&M&N&L\\
\cline{1-1}
IV at $T \ll T_{c}$&$1.9 \times 10^{7}$&$2.0 \times 10^{7}$&
$2.3 \times 10^{7}$&$1.7 \times 10^{7}$\\
R(T) shift at $T \sim$ \tcy &$2.2 \times 10^{7}$&$2.1 \times 10^{7}$
&$2.0 \times 10^{7}$&$1.8 \times 10^{7}$\\
\hline \hline
\end{tabular}

\vspace{1em}

\begin{tabular}{||l|r|r|r|r||}
\hline \hline
Theory:&\multicolumn{4}{|c||}{}\\
\cline{1-1}
GL&\multicolumn{4}{|c||}{$4 \times 10^{7}$}\\
QPE shift&\multicolumn{4}{|c||}{$7 \times 10^{7}$}\\
\hline \hline
\end{tabular}
\caption{Zero-temperature values of \jd in A/\cmsy }
\end{table}
These are seen to be consistent with the values (bottom row of numbers
in Table~1) obtained from shifts in the resistive
transitions near \tc (Fig.~\ref{laws} and Eq.~\ref{tcjsmall}).  
It may be reiterated that the measurement does not reflect a
depinning \jcy ; without significant pairbreaking or runaway heating, the
motion of the minuscule self flux ($B \ll $\hcuy ) will produce $R_{f} \ll
R_{n}$. The observed $R\approx R_{n}$ is reached only when the 
superconductivity is completely destroyed and the system has become normal.

The average value for all samples by both methods is \jdy $(0) \approx 
2 \pm 0.7 \times 10^{7}$ A/\cmsy .

Our experimental estimate of \jdy $(0)$ can be compared with
theoretically calculated values from \eqr {jdzeroGL} and \eqr
{nsvs-jd0}. The \hcu that enters these equations is the one that
reflects the clean-limit BCS coherence length in the {\em ab}
crystalline plane ($H \parallel c$ axis). An actual measured \hcu
reflects the reduced coherence length due to scattering 
($\xi \approx \sqrt{\xi_{0}l}$). 
Hence \jd
calculated from an empirically measured \hcu is going to be an overestimate. We
take \hcuy $(0) \approx 3$ T obtained from crystals by Sologubenko 
et al. \cite{ott} as an upper bound on the clean-limit 
\hcuy . Because of sample to sample variation and 
different amounts of impurity scattering, the uncertainty and range in 
measured values of \hcuy, span a factor of
five \cite{other-hcu}. For $\lambda$ we take the value of 150 nm found using
an AC induction technique on the same type of film samples by
Kim et al. \cite{kim}. The experimental uncertainty and range 
in $\lambda$ is about $\pm 40$\% \cite{other-lambda}, which is a tighter
error bar than that of \hcuy .  With these values of
\hcu and $\lambda$ we obtain \jdy$(0)$ based on 
the GL calculation (\eqr {jdzeroGL}) 
 and 
the quasiparticle-energy shift (\eqr {nsvs-jd0}).
These calculated values are also shown in 
Table~1 and are indeed higher than the measured \jd (by about a factor 
of 2--3), but are of the same order of magnitude, which is about the
best agreement that can be expected given the uncertainties in
the parameters. 

Incidentally, from the measured value \cite{kim} of 150
nm and \eqr{lambda}, the superfluid density turns out to be
$n_{s}(0) \approx 1.3 \times 10^{21}$ cm$^{-3}$ and the measured \cite{ott}
\hcuy $(0)
\approx 3$ T and \eqr {hc-hc2-lambda} give $H_{c} \approx 0.18$ T. 

\subsection{Temperature dependence of \idy }
\begin{figure}[h] 
 \begin{center}
	\leavevmode
	\epsfxsize=0.8\hsize
	\epsfbox{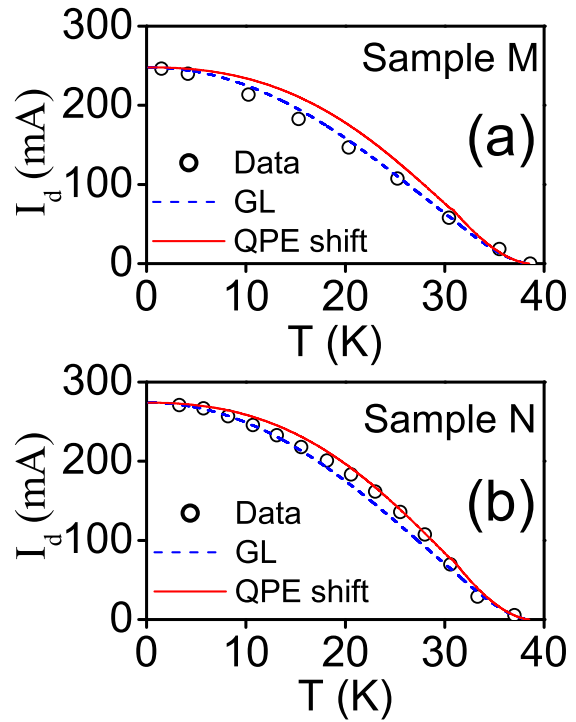}
 \end{center}
 \vspace{-1em}
\caption{{
Pair-breaking currents from IV characteristics at fixed temperatures 
(jumps in Fig.~\ref{ivcurves}). The dashed line represents the GL
\idy $(T)$ function (Eq.~\ref{jdtfull}) and the solid line represents the
theoretical \idy $(T)$ resulting from quasiparticle-energy shift 
(\eqr {nsvs-jdt}). 
The end points $I_{d}(0)$ and \tcuy $(0)$ were fixed by the data; no
other adjustable parameters are involved in the interpolation.}}
\label{glcurve}
\end{figure}
We now take the \id values determined by the $IV$ jumps of \figr{ivcurves}
and plot them versus temperature.  Fig.~\ref{glcurve} 
shows such plots for two samples.
The lines show the theoretical \tdepi expected from the GL treatment
(\eqr {SBGL-jdt1}) and from QPE shifts (\eqr {nsvs-jdt}). For
$H_{c}(T)$ we take the generic empirical temperature dependence 
$H_{c}(T) \approx H_{c}(0)[1-(T/T_{c})^{2}]$, for $\Delta (T)$ we take the
BCS function, and for $\lambda(T)$ we take the empirical \tdep
measured for \mgb by Kim et al. \footnote{For our purpose and 
temperature range of interest ($0 \alt T \alt 0.95 T_{c}$), Kim et al.'s
empirical \tdep \cite{kim} of $\lambda$ can be sufficiently well approximated
by the simple function $\lambda (T) \approx 
\lambda (0)[1-(T/T_{c})^{2.2}]^{-1/2}$ for $T < 0.8$\tc and 
$\lambda (T) \approx 2.74\lambda (0)[1-(T/T_{c})^{1.2}]^{-1/2}$ for $T >
0.8$\tcy . Similarly the BCS \tdep of the gap can be 
approximated by $\Delta (T) \approx \Delta (0)[1 - 
\tan(0.67396 \times (T/T_{c})^{3.7})]$.}.
The ends of the curves [0,\idy $(0)$]
and [\tcuy ($0$),0] are fixed. Other than that there are no adjustable
parameters. As can be seen the data follow the general trend of the theoretical
curves. 

\section{Conclusions}
In conclusion, we have studied current induced 
pair-breaking in magnesium diboride over the
entire temperature range for in-plane current transport. 
The measured \idy $(T)$ function is 
consistent with the expected theoretical \tdepi and conforms exactly 
to the $\Delta T_{c} \propto j^{2/3}$ behavior predicted near \tcy . 
\jdy $(0)$ obtained from the value of current required to drive the 
sample normal at  $T \rightarrow 0$, agrees with the \jdy $(0)$ deduced from
the  $\Delta T_{c} \propto j^{2/3}$ behavior close to \tcy .
The average value from our measurement is \jdy $(0) \approx 
2 \pm 0.7 \times 10^{7}$ A/\cmsy , which is about a factor of three
lower than the value $\sim 6 \times 10^{7}$ A/\cms calculated from
published parameters, but the two are consistent within the
uncertainities of the various parameters.  

From a technological
standpoint, the depairing current density of \mgb is about an order
of magnitude lower than the high-\tc cuprates \cite{pair}. The good news
is that flux motion in films is so quenched \cite{absence} 
that the depinning \jc 
at modest fields appears to be over 25\% the magnitude of
\jd \cite{otherhighjs,otherhighjs2,otherhighjs3}, whereas for the 
cuprates, \jc and \jd can be 
separated by two or three orders of magnitude \cite{mplb}. 

The tremendous experimental difficulties against measuring
\jdy$(0)$ until now, can be appreciated when one sees that
for \ybco (where \jdy$(T\approx T_{c})$ was
measured \cite{pair}) the power density would
be \cite{metal,pair} $\rho j^{2} \sim 10^{-4} (10^{8})^2 \sim 10^{12}$
W/\cmcy!---hopelessly beyond our pulsed technique's limit of
$\sim$$10^{10}$ W/\cmcy . Low-\tc materials like Nb and Pb also have
prohibitive  $\rho j^{2}$ values. \mgby 's parameters
($\rho j^{2} \sim 10^{-5} (10^{7})^2 \sim 10^{9}$ W/\cmcy) 
brought \jdy$(0)$ within experimental reach.

\section{Acknowledgements}
The author acknowledges useful discussions and other assistance from
S. I. Lee, W. Kang, A. Gurevich, D. H. Arcos, G. Saracila, J. M. Knight,
R. Prozorov, B. I. Ivlev, K. Gray, D. Larbalestier, and D. K. Finnemore. 
This work was supported by the U. S. Department of Energy through grant 
number DE-FG02-99ER45763.

\end{document}

%% file: defs.tex
\newcommand{\rb}{\right)}
\newcommand{\lb}{\left(}
\newcommand{\eqr}{Eq.~\ref}
\newcommand{\figr}{Fig.~\ref}
\newcommand{\tdep}{temperature dependence }
\newcommand{\tdepy}{temperature dependence}
\newcommand{\tdepi}{temperature dependencies }
\newcommand{\tdepiy}{temperature dependencies}
\newcommand{\psy}{$\psi$}
\newcommand{\psa}{$|\psi|$ }
\newcommand{\psay}{$|\psi|$}
\newcommand{\del}{$\Delta$ }
\newcommand{\dely}{$\Delta$}
\newcommand{\gt}{$G(T)$ }
\newcommand{\gty}{$G(T)$}
\newcommand{\bb}{$B$ }
\newcommand{\bby}{$B$}
\newcommand{\rth}{$R_{th}$ }
\newcommand{\rthy}{$R_{th}$}
\newcommand{\cll}{$c_{11}$ }
\newcommand{\clly}{$c_{11}$}
\newcommand{\cuu}{$c_{66}$ }
\newcommand{\cuuy}{$c_{66}$}
\newcommand{\cms}{cm$^{2}$ }
\newcommand{\cmsy}{cm$^{2}$}
\newcommand{\cmc}{cm$^{3}$ }
\newcommand{\cmcy}{cm$^{3}$}
\newcommand{\ab}{$\sim$ }
\newcommand{\aby}{$\sim$}
\newcommand{\tp}{$T'$ }
\newcommand{\tpy}{$T'$}
\newcommand{\tph}{$T_{p}$ }
\newcommand{\tphy}{$T_{p}$}
\newcommand{\too}{$T_{0}$ }
\newcommand{\tooy}{$T_{0}$}
\newcommand{\tauep}{$\tau_{ep}$ }
\newcommand{\tauepy}{$\tau_{ep}$}
\newcommand{\tauee}{$\tau_{ee}$ }
\newcommand{\taueey}{$\tau_{ee}$}
\newcommand{\jphi}{$j_{\phi}$ }
\newcommand{\jphiy}{$j_{\phi}$}
\newcommand{\tc}{$T_{c}$ }
\newcommand{\tcy}{$T_{c}$}
\newcommand{\tcu}{$T_{c2}$ }
\newcommand{\tcuy}{$T_{c2}$}
\newcommand{\hcl}{$H_{c1}$ }
\newcommand{\hcly}{$H_{c1}$}
\newcommand{\ef}{$E_{F}$ }
\newcommand{\efy}{$E_{F}$}
\newcommand{\estar}{$E^{*}$ }
\newcommand{\estary}{$E^{*}$}
\newcommand{\htc}{high-temperature superconductors } 
\newcommand{\htcy}{high-temperature superconductors}
\newcommand{\et}{{\it et al. }}
\newcommand{\ety}{{\it et al.}}
\newcommand{\be}{\begin{equation} }
\newcommand{\ene}{\end{equation}}
\newcommand{\hh}{$H$ }
\newcommand{\hhy}{$H$}
\newcommand{\hc}{$H_{c}$ }
\newcommand{\hcy}{$H_{c}$}
\newcommand{\ho}{$H_{0}$ }
\newcommand{\hoy}{$H_{0}$}
\newcommand{\jc}{$j_{c}$ }
\newcommand{\jcy}{$j_{c}$}
\newcommand{\ic}{$I_{c}$ }
\newcommand{\icy}{$I_{c}$}
\newcommand{\sg}{superconducting }
\newcommand{\sgy}{superconducting}
\newcommand{\ssc}{superconductor }
\newcommand{\sscy}{superconductor}
\newcommand{\hcu}{$H_{c2}$ }
\newcommand{\hcuy}{$H_{c2}$}
\newcommand{\rfff}{$\rho_{f}$ }
\newcommand{\rfffy}{$\rho_{f}$}
\newcommand{\hcut}{$H_{c2}(T)$ }
\newcommand{\hcuty}{$H_{c2}(T)$}
\newcommand{\jd}{$j_{d}$ }
\newcommand{\jdy}{$j_{d}$}
\newcommand{\id}{$I_{d}$ }
\newcommand{\jdt}{$j_{d}(T)$ }
\newcommand{\jdty}{$j_{d}(T)$}
\newcommand{\idy}{$I_{d}$}
\newcommand{\ybco}{Y$_{1}$Ba$_{2}$Cu$_{3}$O$_{7}$ }
\newcommand{\ybcoy}{Y$_{1}$Ba$_{2}$Cu$_{3}$O$_{7}$}
\newcommand{\lsco}{La$_{2-x}$Sr$_{x}$CuO$_{4}$ }
\newcommand{\lscoy}{La$_{2-x}$Sr$_{x}$CuO$_{4}$}
\newcommand{\mgb}{MgB$_{2}$ }
\newcommand{\mgby}{MgB$_{2}$}
\newcommand{\de}{$\delta \epsilon$ }
\newcommand{\dey}{$\delta \epsilon$}
\newcommand{\nq}{$n_{q}$ }
\newcommand{\nqy}{$n_{q}$}
\newcommand{\rrhon}{$\rho_{n}$ }
\newcommand{\rrhony}{$\rho_{n}$}
\newcommand{\rrho}{{$\rho$} }
\newcommand{\rrhoy}{{$\rho$}}
\newcommand{\qp}{quasiparticle }
\newcommand{\qpy}{quasiparticle}
\newcommand{\qps}{quasiparticles }
\newcommand{\qpsy}{quasiparticles}
\newcommand{\bib}{\bibitem}
\newcommand{\ib}{{\em ibid. }}
\newcommand{\taue}{$\tau_{\epsilon}$ }
\newcommand{\tauey}{$\tau_{\epsilon}$}
\newcommand{\vstary}{$v^{*}$}
\newcommand{\vstar}{$v^{*}$ }
\newcommand{\tstar}{$T^{*}$ }
\newcommand{\tstary}{$T^{*}$}
\newcommand{\rhostar}{$\rho^{*}$ }
\newcommand{\rhostary}{$\rho^{*}$}
\newcommand{\vinf}{$v_{\infty}$ }
\newcommand{\vinfy}{$v_{\infty}$}
\newcommand{\fd}{$F_{d}$ }
\newcommand{\fdy}{$F_{d}$}
\newcommand{\fe}{$F_{e}$ }
\newcommand{\fey}{$F_{e}$}
\newcommand{\fl}{$F_{L}$ }
\newcommand{\fly}{$F_{L}$}
\newcommand{\jstar}{$j^{*}$ }
\newcommand{\jstary}{$j^{*}$}
\newcommand{\ns}{$n_{s}$ }
\newcommand{\nsy}{$n_{s}$}
\newcommand{\vs}{$v_{s}$ }
\newcommand{\vsy}{$v_{s}$}
\newcommand{\je}{$j(E)$ }
\newcommand{\jey}{$j(E)$}
\newcommand{\vphi}{$v_{\phi}$ }
\newcommand{\vphiy}{$v_{\phi}$}
\newcommand{\blo}{$B_{1}$ }
\newcommand{\bloy}{$B_{1}$}
\newcommand{\bhi}{$B_{\infty}$ }
\newcommand{\bhiy}{$B_{\infty}$}
\newcommand{\vlo}{$v_{1}$ }
\newcommand{\vloy}{$v_{1}$}
\newcommand{\bo}{$B_{o}$ }
\newcommand{\boy}{$B_{o}$}
\newcommand{\eo}{$E_{o}$ }
\newcommand{\eoy}{$E_{o}$}